\documentclass[a4paper,10pt,onecolumn,preprint,sort&compress]{elsarticle}
\usepackage[english]{babel}
\usepackage{amsmath,amsthm,amsfonts,amssymb,natbib}
\usepackage{dsfont}
\usepackage{xcolor}
\usepackage{hyperref}
\journal{Physics Letters B}
\DeclareMathOperator{\e}{e}
\DeclareMathOperator{\sech}{sech}

\graphicspath{{./img/}}

\begin{document}

\begin{frontmatter}

\title{Phase transitions in thick branes endorsed by entropic information}

\author[ifce]{W. T. Cruz}
\ead{wilamicruz@gmail.com}

\author[pici]{D. M. Dantas}
\ead{davi@fisica.ufc.br }

\author[unesp,ita]{R. A. C. Correa}
\ead{rafael.couceiro@ufabc.edu.br}

\author[pici]{C. A. S. Almeida}
\ead{carlos@fisica.ufc.br}

\address[ifce]{Instituto Federal de Educa\c{c}\~ao, Ci\^encia e Tecnologia do Cear\'a (IFCE), Campus
Juazeiro do Norte - 63040-540 Juazeiro do Norte, CE, Brazil}

\address[pici]{Universidade Federal do Cear\'a (UFC), Departamento de F\'isica, Campus do Pici, Fortaleza - CE, C.P. 6030, 60455-760 - Brazil}

\address[unesp]{UNESP, Universidade Estadual Paulista, Campus de Guaratinguet\'a, 12516-410,
Guaratinguet\'a, S\~ao Paulo, Brazil}

\address[ita]{ITA, Instituto Tecnol\'ogico de Aeron\'autica- Departamento de F\'isica,
12228-900, S\~ao Jos\'e dos Campos, S\~ao Paulo, Brazil}

\begin{keyword}
Bloch brane \sep brane-world models \sep configurational entropy
\end{keyword}

\begin{abstract}
The so-called configurational entropy (CE) framework has proved to be an efficient instrument to study nonlinear scalar field models featuring solutions with spatially-localized energy, since its proposal by Gleiser and Stamapoulos. Therefore, in this work, we apply this new physical quantity in order to investigate the properties of degenerate Bloch branes. We show that it is possible to
construct a configurational entropy measure in functional space from the
field configurations, where a complete set of exact solutions for the model
studied displays both double and single-kink configurations. Our study shows
a rich internal structure of the configurations, where we observe that the
field configurations undergo a quick phase transition, which is endorsed by
information entropy. Furthermore, the Bloch configurational entropy is
employed to demonstrate a high organisational degree in the structure of the
configurations of the system, stating that there is a best ordering for the
solutions.
\end{abstract}
\end{frontmatter}

\section{INTRODUCTION}

In the last years, the study of phenomenological properties in braneworld
models are increasing. We have recent works explaining anomalies in the
meson $B$ decay \cite{lhcb} and in the neutrinos physics \cite{valle},
performing bounds into corrections to Coulomb's law \cite{cou1, cou2}, into
electrical conductivity \cite{condu}, as also adjusting parameters to the
experimental data of the nucleon-nucleus total cross-section of various
chemical elements \cite{cross}. Applications to Myers-Perry's black holes and results based on the observation of gravitational waves are others interesting issues about new insights in the braneworld framework.

In this context, a well-known model is the Bloch brane scenario, proposed by
Bazeia and Gomes \cite{bloch_brane}. This thick brane model is generated by
two interacting scalar fields that  perform the thickness of the
model that solves several issues in the fields localisation present in thin
models \cite{5Dthick, Bajc}, as well as in the Randall-Sundrum models \cite{RS1,RS2}.
Moreover, the structure of this scenario is based on domain walls, which
have some interesting application in several branches of Physics as in high
energy physics \cite{vilenkin}, cosmology \cite{cosm, cosm2}, quantum field
theory \cite{qft} and propositions about the  gravitational waves
observation \cite{gravw}.
%can be assimilated in supergravity \cite%{degenerate-Bloch-Brane-1}. These two scalar fields

The authors of the works \cite{PLBdutra, degenerate-Bloch-Brane-1,
degenerate-Bloch-Brane-2} have shown the existence of more
general Bloch brane solutions in which a degeneracy parameter is responsible
for the raising of two-kink solutions. That model is the so-called degenerate Bloch brane (DBB), where the energy of the field configuration in the
superpotential is precisely the same with regarding all
parameter associated with the domain wall \cite{degenerate-Bloch-Brane-1}. So, we have the
formation of a double brane structure with a splitting effect that is
magnified by the approaching of the degeneracy parameter to a critical
value. 

This transition from single-kink solution to double-kink (or
multi-kink) solutions have some physical implications. The multi-kinks appear in dispersive non-linear system, where single-kinks are no more stables \cite{m1}. The Ref. \cite{m4} shows the appearance of double-kink soliton in the sine-Gordon model under the perturbation of a space-dependent force.  The experimental application of multi-kink concepts arises, among other, in the mobility hysteresis in a damped driven commensurable chain of atoms \cite{m2} and in arrays
of Josephson junctions \cite{m3}. 
Specifically in the context of  braneworlds, the presence of the massive resonant Kaluza-Klein modes is dependent upon transition parameter. A resonant KK mode is an extradimensional massive mode of a field with a finite lifetime, which can bring some interesting phenomenology to braneworlds branch \cite{lhcb,cou2,rizzo, photon1, julio2}. The study of resonances for gravity and fermions in the symmetric and asymmetric cases of the usual Bloch branes was performed in Ref. \cite{Bloch-fg}. However,  more interesting results are present in the DBB scenarios \cite{PLBcruz2014, degenerate-Bloch-Brane-2}. In this case the degeneracy constant tends to a critical value,  a highly KK gravity mode
coupled to the brane is observed \cite{PLBcruz2014}, and also crucial issues in the localisation of massless fermions are clarified \cite{degenerate-Bloch-Brane-2}. Moreover, the gauge field resonances are present only in the double-kink version of sine-Gordon models \cite{cou2}. In short, the double-kink models bring more stability to the models and richer physical applications.
% Lastly, such a picture allows a deeper study of first-order phase transitions in the context of brane worlds to describe brane splitting, which is a fundamental ingredient for the brane cosmological models.
 
%, where the informational contents are studied in models with localised energy configuration.
On the other hand, the reference \cite{PLBgleiser-stamatopoulos}
reintroduces the concept of informational entropy, based on Shannon's
information entropy \cite{shannon}. The so-called Configurational
Entropy (CE) was constructed and applied to several nonlinear scalar field
models featuring solutions with spatially-localised energy. As presented in
Ref. \cite{PLBgleiser-stamatopoulos}, the CE can solve energies of
degenerate configurations. The approach presented in \cite%
{PLBgleiser-stamatopoulos} have been used to study the non-equilibrium
dynamics of spontaneous symmetry breaking \cite{PRDgleiser-stamatopoulos},
to obtain the stability bound for compact objects like Q-balls \cite%
{PLBgleiser-sowinski}, to investigate the emergence of localized objects
during inflationary preheating \cite{PRDgleiser-graham}, and moreover to
distinguish configurations with energy-degenerate spatial profiles
\cite{Rafael-Dutra-Gleiser}. The CE bounds the stability of various
self-gravitating astrophysical objects \cite{Gleiser-Jiang} and states in
Lorentz Violating (LV) scenarios \cite{Rafael-Roldao} and also provides
information about the stability of the glueball states in a dynamical
holographic AdS/QCD model \cite{glueball}. In the topic of braneworlds, the
CE was heretofore applied to sine-Gordon models \cite{bc}, to models with $%
f(R)$ \cite{Rafael-Pedro}, to $f(R,T)$ \cite{Rafael-Pedro2} theories of
gravity, and to the Weyl brane \cite{weylb}, as well as to the topological
abelian string-vortex in six dimensions \cite{6dce}.

Guided by the previous results involving degenerate two-field thick brane
solutions, we propose in this work to investigate the Bloch brane solutions
and degenerate versions by means of the CE information.

%We start the investigation from the energy density along the extra dimension of such brane models.

%This work is organized as follows: In the next section we introduce the
%necessary background for the present study. In the sequel, we formulate a
%configurational entropy measure in the functional space, from the field
%configurations where the Bloch brane can be studied. We conclude with some
%general remarks on our results and directions for future works.

\section{\textbf{Bloch brane overview}}

% , such as those involving the cosmological constant, hierarchy, flavor problem, grand unification, and the origin of the eletroweak symmetry breaking or supersymmetry breaking

%Due to the warp
%geometry of the spacetime, the braneworld scenarios can provide, in a
%straightforward and clear manner, the localizing of a massless graviton on
%the brane and then reproducing effectively four-dimensional gravity at large
%distances. For this and many other reasons that braneworld theories has
%opened new directions for tackling outstanding questions in physics.}

%An interesting consequence from thick branes, it was observed by Campos a few years ago. In that work,
%The genealogy of this name come from condensed
%matter systems, where ferromagnetic materials display domain walls with a
%prominent internal structure, which may be seen as chiral interfaces between
%domains with different magnetization.

%\textbf{improve}
%{In this section, we will show an overview about the so-called
%bloch branes \cite{bloch_brane}.
%we can obtain solutions to Einstein's equation coupled to a single scalar field. In this context,
A very interesting class of configuration in theories involving
extra dimensions is that one where the scalar field
give rise a domain wall, which is baptised in the literature as thick brane. It was shown that some kinds of two interacting
scalar field potentials can be used in order to describe the splitting of
thick branes due to a first-order phase transition in a warped geometry. As
a consequence, we can find remarkable and distinctive critical phenomena in
warped spacetimes, which can open a new window to study cosmological
scenarios. Other efficient alternative to find answers for the cosmological
issues come from the work by Bazeia and Campos{\ \cite{bloch_brane}}, where
it was studied a system described by two real scalar fields coupled with
gravity in $\left(4+1\right)$ dimensions in warped spacetime involving one extra
dimension. In that work, it was found a rich class of brane configuration,
which was called Bloch brane. The most important feature of the
Bloch brane solutions is its stability regarding the classical linear
fluctuations of the scalar fields.
%
% Furthermore, the presence of nontrivial
%structure inside the brane enables to control the energy density in order to
%localise inside the brane.

%Therefore, in order to begin our investigation on configurational entropy in
%Bloch brane backgrounds, in the subsections that follows, we will present
%the warped five-dimensional setup with two interacting scalar fields. From
%specific scalar field potentials, we achieve first order solutions to the
%equations of motion and thereby the warp factor.

%As discussed earlier, the physics of branes in higher dimensional theories can play an important role for solving a large number of physical problems.

\subsection{\textbf{Bloch brane}}

The simplest Bloch brane setup is built with the coupling of two fields
to gravity, as we describe below. The scalar fields depend only on the extra
dimension $y$. The usual action in five-dimensional ($5D$) gravity can be represented by 
\begin{equation}
S=\int {d^{4}xdy}\sqrt{\lvert g\lvert }\left[ -\frac{R}{4}+\frac{1}{2}\left(
\partial _{\mu }\phi \partial ^{\mu }\phi +\partial _{\mu }\chi \partial
^{\mu }\chi \right) -V\left( \phi ,\chi \right) \right] ,  \label{action}
\end{equation}%
where $g=det(g_{\mu \nu })$ and $R$ is the curvature scalar for the metric $%
ds^{2}=\e^{2A}\eta _{\mu \nu }dx^{\mu }dx^{\nu }-dy^{2}$. From Eq. (\ref{action}), we can obtain the following equation of motion and
the corresponding modified Einstein equations
\begin{gather}
\phi ^{\prime \prime }+4A^{\prime }\phi ^{\prime }=\frac{\partial V}{%
\partial \phi }, \quad
\chi ^{\prime \prime }+4A^{\prime }\chi ^{\prime }=\frac{\partial V}{%
\partial \chi } \\
A^{\prime \prime }=-\frac{2}{3}\left( \phi ^{\prime 2}+\chi ^{\prime
2}\right) \\
A^{\prime 2}=\frac{1}{6}\left( \phi ^{\prime 2}+\chi ^{\prime 2}\right) -%
\frac{1}{3}V(\phi ,\chi ).
\end{gather}
where prime stands for derivative with respect to $y$.

In order to obtain first order equations from the equations of motion, let
us apply the so-called superpotential method \cite{bazeia,
bazeia2, bazeia3, bazeia4, shifman, alonso, bazeia5}
% {\color{red}In this case, the potential $V(\phi ,\chi )$\ can be written in terms of a kind of superpotential, where its representation in given by}%
\begin{equation}
V(\phi ,\chi )=\frac{1}{8}\left[ \left( \frac{\partial W}{\partial \phi }%
\right) ^{2}+\left( \frac{\partial W}{\partial \chi }\right) ^{2}\right] -%
\frac{1}{3}W^{2}(\phi ,\chi ).  \label{potbloch}
\end{equation}%
where $W(\phi ,\chi )$ is the superpotential, which from Ref. \cite{bazeia} is defined by
\begin{equation}
W(\phi ,\chi )=2\phi -\frac{2}{3}\phi ^{3}-2r\phi \chi ^{2}\ ,\ 
\end{equation}%
where $r$ is a real thickness parameter that can vary in the interval $%
r\in (0,1/2)$.

%From now on, as an illustrative example, we will study the model
%introduced in Ref. \cite{bazeia}, which guarantees the stability of any
%finite energy solution of the associated first-order system of equations
%through the classical Bogomol'nyi-Prasad-Sommerfield approach. In this case,
%the superpotential is defined by } 
%, which is a smooth
%function of the fields $\phi $\ and $\chi $\ at each point in $R^{2}$.

%\textbf{improve}
We want to stress, however, that this particular superpotential
has a fertile structure being very useful in a large number of physical
applications. For instance, studies include topological defects,
localisation of fermions on critical branes, supersymmetric theories,
travelling solitons in Lorentz and CPT breaking systems, bags, junctions, and
networks of BPS and non-BPS defects \cite{bazeia,bloch_brane,PLBdutra,degenerate-Bloch-Brane-1,
degenerate-Bloch-Brane-2}.

With the potential introduced in Eq.(\ref{potbloch}), we obtain the
resulting first-order equations $\phi ^{\prime }=\frac{1}{2}\frac{\partial W%
}{\partial \phi }$, $\chi ^{\prime }=\frac{1}{2}\frac{\partial W}{\partial
\chi }$ and $A^{\prime }=-\frac{1}{3}W$, from which we find the solutions
that describe our brane model. The solutions to the fields are %\label{sol2}
\begin{gather}
\phi (y) =\tanh (2ry),\ \chi (y) =\pm \left( \sqrt{\frac{1}{r}-2}\right) \sech(2ry)\ .
\label{sol1}
\end{gather}
%It is important to remark that the above solutions also solve the equations of motion. 
The upper limit $r\rightarrow 1/2$
changes the Bloch brane profile, where two-field solution turns into
one-field solution \cite{bloch_brane}. However, for the usual Bloch brane
the two-kink profile is never achieved by variations in $r$. The warp
factor $e^{2A(y)}$ is obtained by the solution of $%
A^{\prime }=-W/3$ in the form
\begin{equation}
A(y)=\frac{1}{9r}\left[ (1-3r)\tanh ^{2}(2ry)-2\ln \cosh (2ry)\right] .
\label{wfunc}
\end{equation}%
For this model, the energy density is written as \cite{bloch_brane} % static configurations of two interacting real scalar fields coupled to gravity 
\begin{equation}
\varepsilon (y)=\e^{2A(y)}\left[ \frac{1}{2}\phi ^{\prime 2}+\frac{1}{2}\chi
^{\prime 2}+V(\phi ,\chi )\right] .  \label{en}
\end{equation}

\begin{figure}[tbh]
\begin{center}
%\textbf{\includegraphics[width=.4\textwidth]{CE1.pdf} }
\textbf{\includegraphics[width=.8\textwidth]{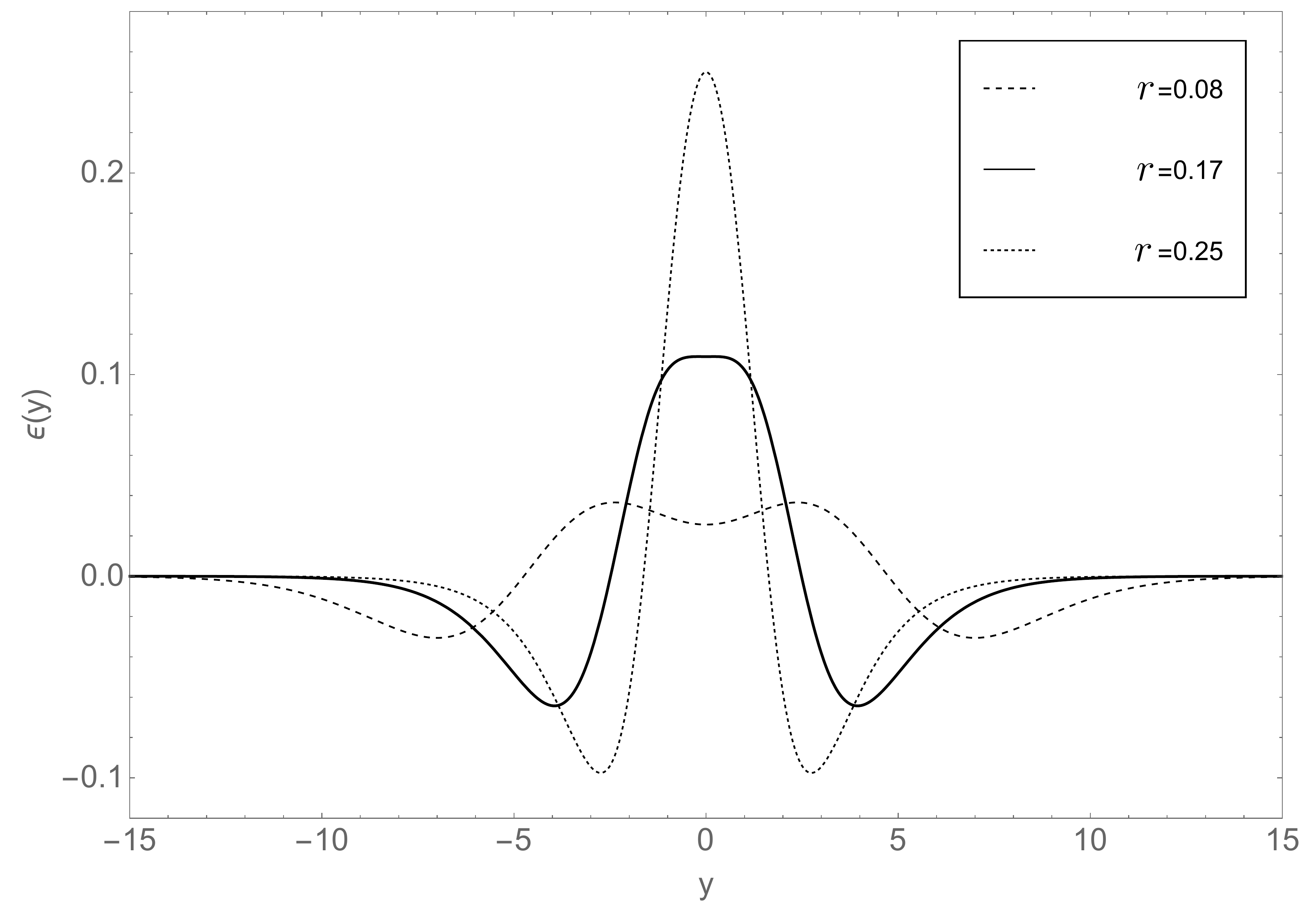} }
\end{center}
\caption{$\varepsilon (y)$ for Bloch brane.}
\label{CE1}
\end{figure}

We plot $\varepsilon (y)$ in Fig. (\ref{CE1}), that show us the appearance
of two peaks on the energy density for the interval $0>r>0.17$. In Ref. \cite{bloch_brane} the authors have noticed the existence of a
brane internal structure that is suppressed by the presence of gravity. The
raising of such structure is related with a specific value of $r$. This
issue will be addressed in the next section by CE concepts.

% where we will analyze the Bloch brane structure as well as the compatibility of these internal structures
%from the point of view of configurational entropy information. 
%{\color{red} In the next subsection, we will show another class of Bloch brane, which is know in the literature as degenerate Bloch brane (DBB).}

\subsection{\textbf{Degenerate I Bloch Brane}}

%\textbf{improve}
%{As mentioned previously, works reporting the appearance of Bloch
%branes with internal structure has appeared in the literature. 
Among other
types of nonlinear field configurations coupled to gravity in (4 + 1)
dimensions in warped space-time with one extra dimension. That is a
specially important class of Bloch branes, which was coined degenerate Bloch
brane (DBB) \cite{degenerate-Bloch-Brane-1, degenerate-Bloch-Brane-2}, due to degenerate energy in the superpotencial parameters. These
new class of configurations, on the contrary of the usual Bloch branes,
enable a control over the brane thickness without needing to change the
potential parameter, but by means of a domain wall degeneracy parameter.
%. In this case, this is done by means of

 In such models, the scalar field manifests a transition from kink to double
kink solution when a degeneracy parameter reaches a critical value with the following potential: 
\begin{equation}
V(\phi ,\chi )=\frac{1}{2}\left[ \left( \frac{\partial W}{\partial \phi }%
\right) ^{2}+\left( \frac{\partial W}{\partial \chi }\right) ^{2}\right] -%
\frac{4}{3}W^{2}(\phi ,\chi ),  \label{w12}
\end{equation}%
%
%
%
%
%
%
%
%
%
%
%
%
%
%
%
%
%
%
%
%
%\textcolor{gray}{}
%{\color{red}
%Now, we will be studying a very useful superpotential, which was
%used for modeling a great number of systems, which is given by 
and the new superpotential
\begin{equation}
W(\phi ,\chi )=\phi \left[ \lambda \left( \frac{\phi ^{2}}{3}-a^{2}\right)
+\mu \chi ^{2}\right],  \label{w13}
\end{equation}
where $\lambda$, $a$ and $\mu$ are real parameters that deform this superpotential.
% It has been shown in Ref. \cite{PLBdutra} that general solutions of the first-order differential equations for the scalar fields can be obtained. In that case, 
 
So, it has been found in Ref. 
\cite{PLBdutra} that there are two particular cases where the first-order
differential equations can be solved analytically. The first set of
solutions is given by \cite{degenerate-Bloch-Brane-1}
\begin{eqnarray}
\chi ^{(1)}(y) &=&\frac{2a^{2}}{\left( \sqrt{c_{0}^{2}-4a^{2}}\right) \cosh
(2\mu ay)-c_{0}},\quad  \label{sol3} \\
\phi ^{(1)}(y) &=&\frac{a\left( \sqrt{c_{0}^{2}-4a^{2}}\right) \sinh (2\mu
ay)}{\left( \sqrt{c_{0}^{2}-4a^{2}}\right) \cosh (2\mu ay)-c_{0}},
\end{eqnarray}%
where $c_0$ is the degeneracy parameter that also regulates the brane thickness (the larger is $c_0$, the thinner is the brane) and it was considered as $c_{0}<-2a$ and $\lambda =\mu $. 
Moreover, in this case, the corresponding warp factor is 
written in the form

\begin{gather}
e^{A(y)}=N\left[ \frac{2a^{2}}{\left( \sqrt{c_{0}^{2}-4a^{2}}\right) \cosh
(2\mu ay)-c_{0}}\right] ^{\frac{4}{9}a^{2}}\times  \notag \\
\times \exp \left\{ \frac{2a^{2}\left[ c_{0}^{2}\pm 4a^{2}-c_{0}\left( \sqrt{%
c_{0}^{2}-4a^{2}}\right) \cosh (2a\mu y)\right] }{9\left[ \left( \sqrt{%
c_{0}^{2}-4a^{2}}\right) \cosh (2\mu ay)-c_{0}\right] ^{2}}\right\} .
\label{sol3a}
\end{gather}

An interesting feature of these solutions is that, for some
values of $c_{0}$ close to the critical value, namely $c_{0}^{crit}=-2a$,
the scalar field $\phi ^{(1)}(y)$ exhibits a double kink profile that
reflects a formation of a double wall structure, extended along the extra
dimension. On the other hand, the scalar field $\chi ^{(1)}(y)$, close to
the critical value $c_{0}^{crit}$, exhibits a flat top. In addition, we can
observe in the warp factor the emergence of a controllable flat region,
where one could think in a Minkowski-type metric region sandwiched between
the two branes.
 
\begin{figure}[tbh]
\begin{center}
%\textbf{\includegraphics[width=.4\textwidth]{CE2.pdf} }
\textbf{\includegraphics[width=.8\textwidth]{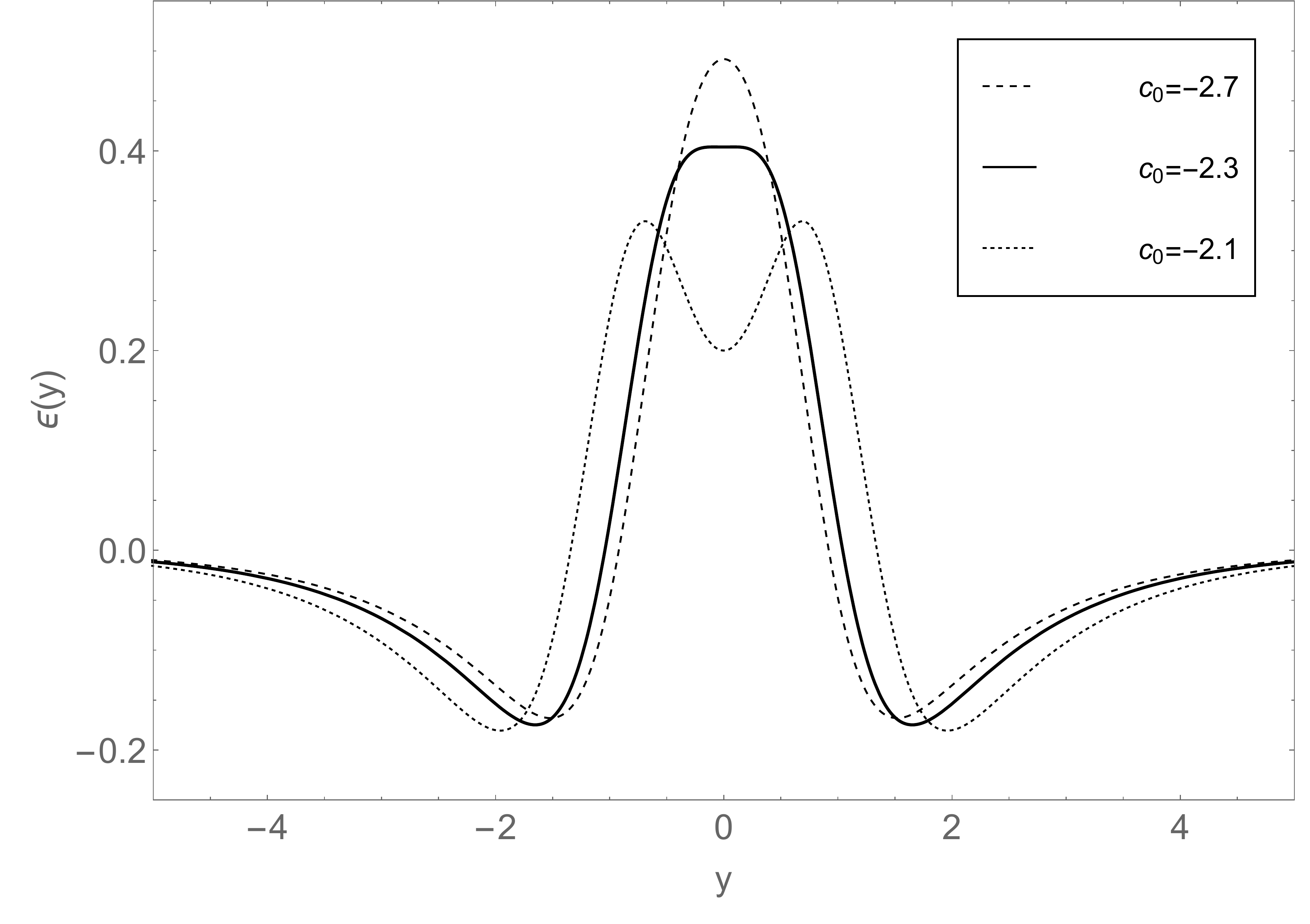} }
\end{center}
\caption{$\protect\varepsilon (y)$ for DBB I with $a=\mu=1$.}
\label{CE2}
\end{figure}

 As we are interested in analysing the information-entropic
measure of these configurations, it is interesting to show the energy
density profile. Therefore with the results above we are able to construct the
energy density (\ref{en}), which is displayed in Fig. (\ref{CE2}%
). From that figure, we can also note the appearance of two
peaks when the degeneracy parameter is close to the critical value,
signalising a richer structure for energy density.

\subsection{\textbf{Degenerate II Bloch Brane}}

Following Ref. \cite{degenerate-Bloch-Brane-2}, with the
same potential \eqref{w12} and the superpotential function \eqref{w13}, 
there is another class of degenerate Bloch brane solution, which we named DBB II. In this case, assuming $c_{0}<1/(16a^{2})$ and $\lambda =4\mu $, we find the following analytical solutions \cite{degenerate-Bloch-Brane-2} 
\begin{equation}
\chi ^{(2)}(y)=-\frac{2a}{\sqrt{\left( \sqrt{1-16c_{0}a^{2}}\right) \cosh
(4\mu ay)+1}},
\end{equation}%
\begin{equation}
\quad \phi ^{(2)}(y)=\frac{a\left( \sqrt{1-16c_{0}a^{2}}\right) \sinh (4\mu
ay)}{\left( \sqrt{1-16c_{0}a^{2}}\right) \cosh (4\mu ay)+1}.
\end{equation}%
Furthermore, the corresponding warp factor is given by 
\begin{gather}
e^{2A(y)}=N\left[ -\frac{2a}{\sqrt{\left( \sqrt{1-16c_{0}a^{2}}\right) \cosh
(4\mu ay)+1}}\right] ^{\frac{16a^{2}}{9}}\times  \notag \\
\times \exp \left\{ -\frac{4a^{2}\left[ 1+8c_{0}a^{2}+\left( \sqrt{%
1-16c_{0}a^{2}}\right) \cosh (4a\mu y)\right] }{9\left[ \left( 1+\sqrt{%
1-16c_{0}a^{2}}\right) \cosh (4\mu ay)\right] ^{2}}\right\} .
\label{sol4a}
\end{gather}

The solution presented in \eqref{sol4a} is also known as critical Bloch brane due the appearance of a more pronounced flat region when the degeneracy parameter  tends to its upper limit $c_0\to\left(1/16a^2\right)$. After all, we note that the DBB I and DBB II are very similar models, as we can see by comparing the warp factors in the equations \eqref{sol3a} and \eqref{sol4a}.

Finally, the resulting energy density is plotted in Fig. (\ref%
{fig-energy-bloch-1}). 
\begin{figure}[tbp]
\begin{center}
%\textbf{\includegraphics[width=.4\textwidth]{CE3.pdf} }
\textbf{\includegraphics[width=.8\textwidth]{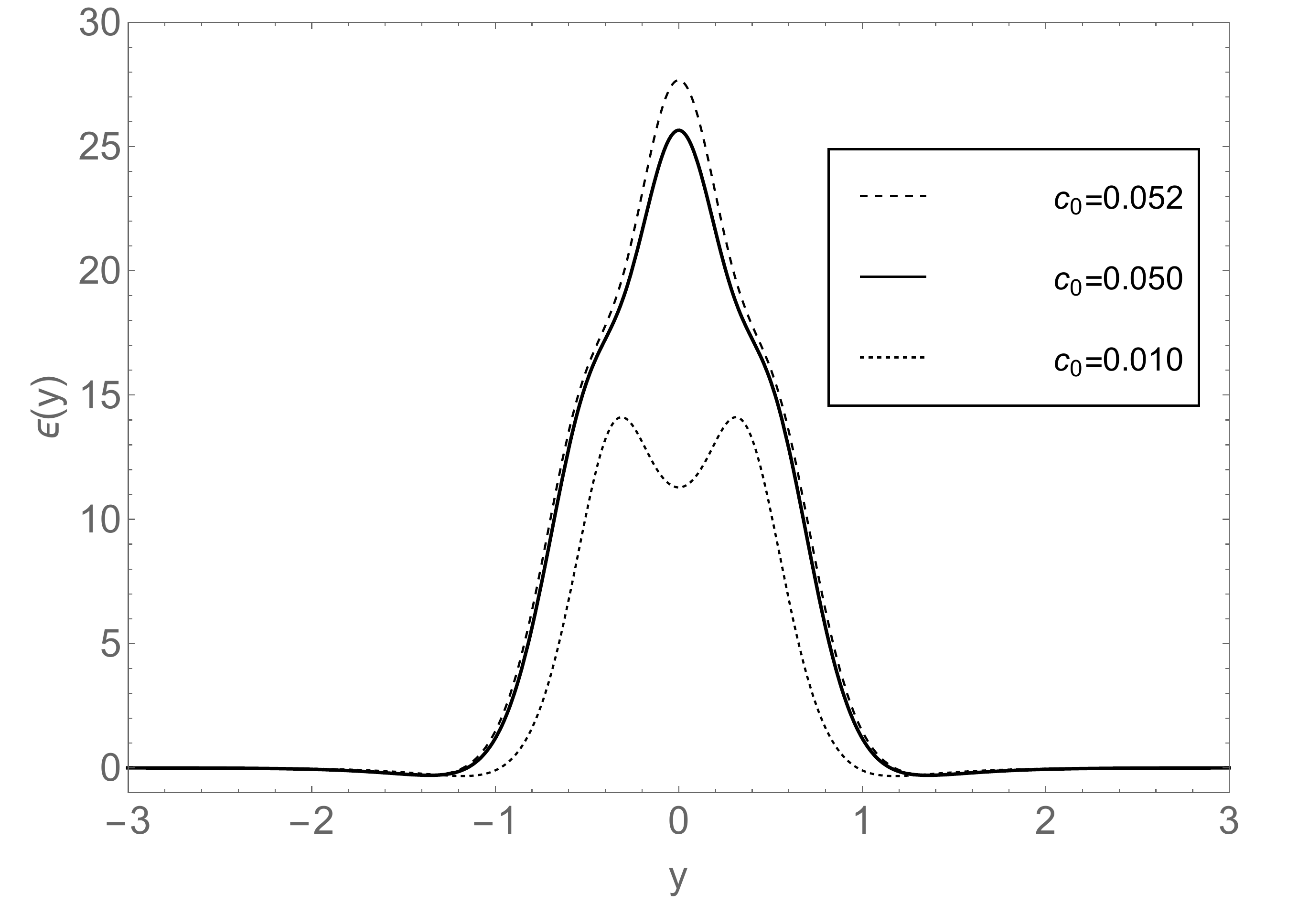} }
\end{center}
\caption{$\protect\varepsilon(y)$ for DBB II with $a=\mu=1$.}
\label{fig-energy-bloch-1}
\end{figure}
The behaviour of the energy density of the DBB II solution is
different from the two previous cases, since its energies are more localised and the peaks are more prominent. However, the DBB II model maintains the splitting of the maxima. In next section, we will verify by CE concepts, that these changes in the DBB II occur because the region where CE tends to the minimum is more narrow that in the DBB I and in the usual Bloch Brane cases.

\section{\textbf{Configurational Entropy in Bloch brane scenario}}

%In this section, we apply an original quantity that can be employed to
%quantify the existence of non-trivial spatially localised solutions in field
%configuration space. 
The so-called configurational entropy (CE) \cite%
{PLBgleiser-stamatopoulos} is correlated to the energy of a localised field
configuration, where low energy systems are linked with small entropic
measures \cite{PLBgleiser-stamatopoulos}.

The CE can be expressed by the following Fourier transform of the energy density \cite{bc, Rafael-Pedro, PLBgleiser-stamatopoulos}

\begin{equation}
\mathcal{F}(\omega )=\frac{1}{\sqrt{2\pi }}\int_{-\infty }^{\infty }{%
\varepsilon (y)\e^{i\omega y}dy}.
\end{equation}

The model where we apply the CE contains structures with spatially localised,
square-integrable, bounded energy density functions $\varepsilon (y)$.
Hence, we can define the so-called modal fraction that reads \cite%
{PLBgleiser-stamatopoulos, PLBgleiser-sowinski, PRDgleiser-stamatopoulos,
Rafael-Dutra-Gleiser} 
\begin{equation}
f(\omega )=\frac{{\lvert \mathcal{F}(\omega )\lvert ^{2}}}{{\int_{-\infty
}^{\infty }{d\omega \lvert \mathcal{F}(\omega )\lvert ^{2}}}\,}.
\end{equation}

Subsequently, we can work with the normalized modal fraction, defined as the
ratio of the normalised Fourier transformed function and its maximum value $\tilde{f}(\omega )={f(\omega )}/{f_{max}} .$ So, localised and continuous function $\tilde{f}(\omega )$ yields the
following definition for the CE
\begin{equation}
S(\tilde{f})=-\int_{-\infty}^{\infty }{d\omega }\tilde{f}(\omega )\mbox{ln}%
\left[ \tilde{f}(\omega )\right] .  \label{CE}
\end{equation}

% $%f_{max}$ 
%\begin{equation}
%\tilde{f}(\omega )={f(\omega )}/{f_{max}} .
%\end{equation}

From the point of view of CE approach, it has been shown that is possible to obtain
important bounds in theories where unknown parameters are presented.  Here, it
should be noted that the strategy of using $c_{0}$ to map a different range of
parameters for the energy density and the CE has been used successfully
several times before. It is worth highlighting that the CE method was already employed
for the flat case of degenerate kinks in two-field models \cite{Rafael-Dutra-Gleiser}.
In the present paper, we verify the influence of the gravity in these degenerate models and
its implications on the fields localisation and the changes in the critical points.
The main motivation in our study is to show that there is a bound regarding the internal structure of
DBB, where the warped geometry leads to the emergence of a controllable flat
region, which is described by a Minkowski space-time. Through this analysis,
we expect to understand the phase transition phenomena in Bloch branes
scenarios, providing tools for a better understanding of new brane
cosmological models. Furthermore, we can point out the most probable parameter that
allows us to have normalisable fermion zero-mode localised on the brane. In addition we have the value for the highly KK gravity modes
coupled in these DBB models. Our
methodology consists in quantify the CE in terms of the
degeneracy parameter $c_{0}$. On this way, we can merge the entropy
information with details of the structure of the defects like thickness and
curvature.
\begin{figure}[tbh]
\begin{center}
%\textbf{\includegraphics[width=.4\textwidth]{CE7.pdf} }
\textbf{\includegraphics[width=.8\textwidth]{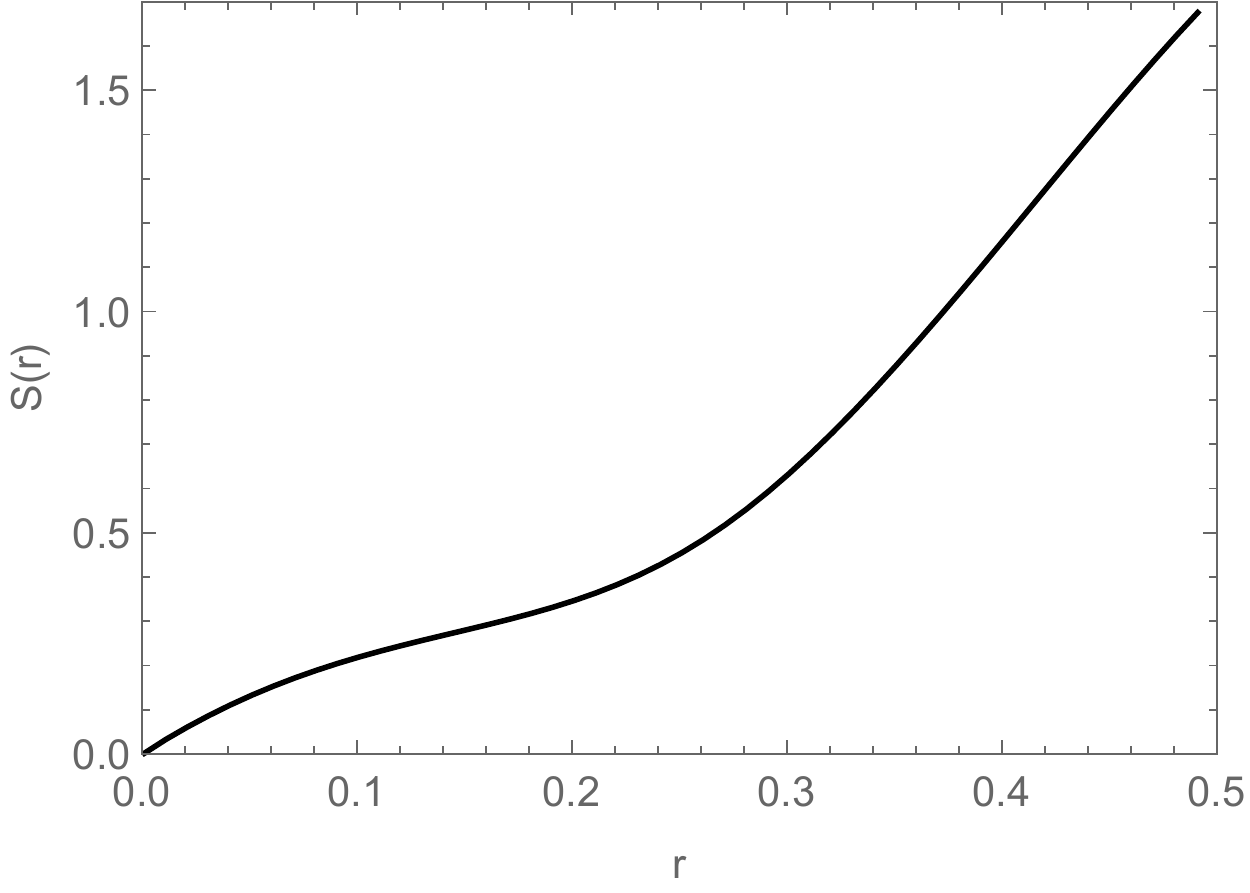} }
\end{center}
\caption{CE for usual Bloch brane solution.}
\label{CE7}
\end{figure}

\subsection{CE in the usual Bloch Brane}
% and (\ref%{sol2})
Now, we review the CE approach to the usual Bloch brane. Firstly we analyse the basic two-field setup from Eqs. (\ref{sol1}). Due to the complexity of the solutions, we evaluate numerically the 
$S(\tilde{f})$ and show the result in Fig. (\ref{CE7}). The graphic shows us
that there is no minimum in the CE to this case and $S(%
\tilde{f})$ is reduced as the thickness of the brane is increased ($r$
goes to zero). This result tells us that the region where we have the opening
of the internal structure, namely, $0<r<0.17$, corresponds to the region of
lower CE. Therefore, in the range of lowest CE, the coupling to gravity does not destroy the presence of internal structure for the Bloch brane. From information-entropic measure point of view, that range matches the values of lower energy of the system, and thus explaining the stability of the configurations. This is an important result from CE background, since that the problem concerning the stability of internal structure had not yet been sufficiently answered in the literature.

The lack of a minimum CE for $r\neq 0$ is related to the absence of a
phase transition in the scalar field solutions. In fact, the solution to $%
\phi (y)$ in this case do not presents the transition kink to double-kink
profile. In addition, the CE shows that the most prominent Bloch
configurations are those where the interaction between the fields is weak. In
this case, we can argue that the CE selects usual Bloch brane models that
in a good approximation has its form dominated by the standard $\phi ^{4}$
model.

\subsection{\textbf{CE in the degenerate I Bloch brane}}

\begin{figure}[!htb]
\begin{center}
\includegraphics[width=.8\textwidth]{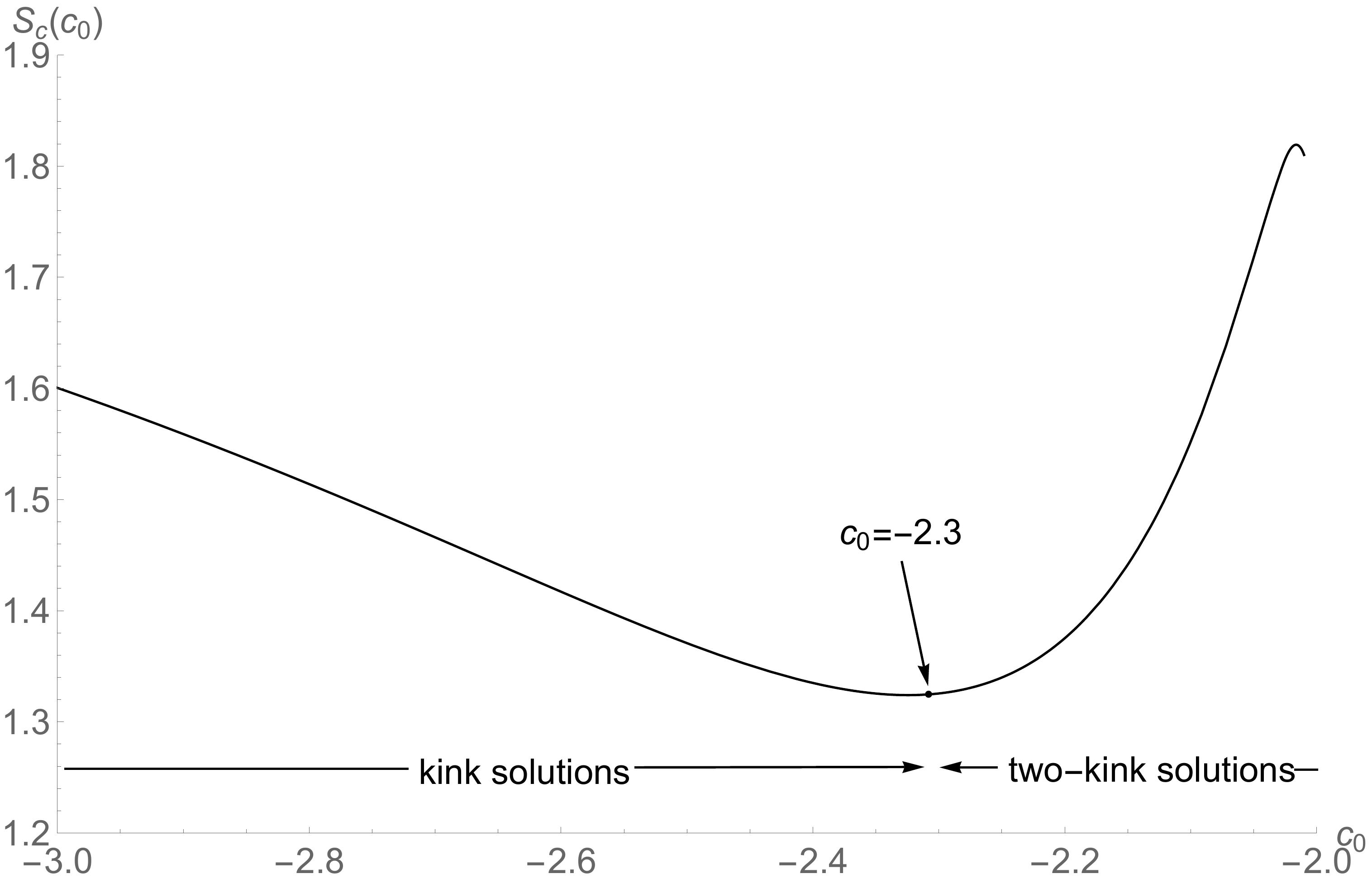}
\end{center}
\caption{CE for DBB I solution with $a=\mu=1$.}
\label{CE9}
\end{figure}

We now turn our attention to the DBB I solution, that is
expressed by the solutions in Eqs. (\ref{sol3}) and (\ref{sol3a}). After
numerical calculations, the CE for this case is presented in Fig. (\ref{CE9}%
). Our fundamental goal is to show that the CE can be used to distinguish between such configurations. Moreover, as a Bloch brane is a thick brane that evolutes to a thicker one, which is called DBB, we are interested in describing the mechanism that can adequately facilitate the description of first order phase transitions. As a matter of fact, these transitions can provide a better understanding of the complex issues regarding brane splitting in warped geometries. In this case, we will see that the CE can be used as a key element for this description.

\begin{figure}[!htb]
\begin{center}
\includegraphics[width=.9\textwidth]{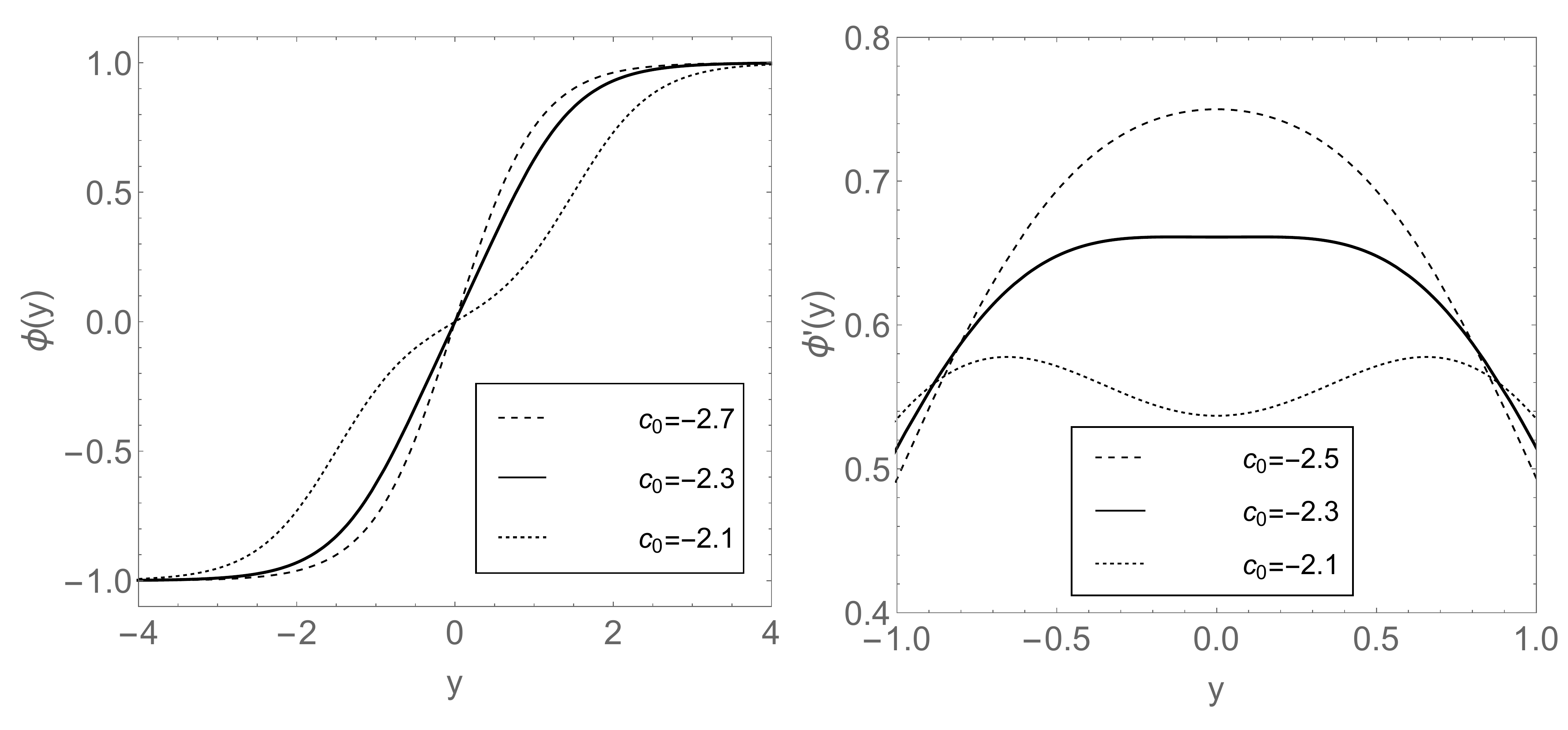}
\end{center}
\caption{Scalar field solution for DBB I case (left) and its first
derivative (right) for $a=\mu=1$.}
\label{CE8}
\end{figure}
From the Fig. (\ref{CE9}), we can note that there is a minimum in the CE at $%
c_0=-2.3$. This point corresponds the phase transition for the scalar field
solution. At this point, we have the raising of two-kink profile. In order
to better identify the raising of the two-kink solutions, we have plotted
the first derivative of the scalar field in Figure (\ref{CE8}). In one-kink
solution, the first derivative must assume a constant value near the
origin. However, when we have the two-kink solution, the $\phi^{\prime }(0)$
must be a local minimum. We note the appearance of a minimum in $%
\phi^{\prime }(0)$ at the critical point ($c_0=-2.3$). Therefore, the
raising of two-kink profile occurs at the degeneracy parameter $c_0$
corresponding to the minimum of the CE.

There is also a correspondence between the entropic information and the
matter-energy density along the extra dimension. For the region where $%
c_0<-2.3$, the $\varepsilon(y)$ degenerate solutions have a single peak
around $y=0$. Our results also show that for $-2.3<c_0<-2.0$ (see fig.( \ref%
{CE2})) the energy density acquires two peaks. The minimum of the
CE at $c_0=-2.3$ is related to the appearance of a
behavior named brane internal structure, as reported in ref. \cite%
{bloch_brane}.

%{\color{red}One very important consequence of our analyze is that}

\subsection{\textbf{CE in the degenerate II Bloch brane}}

\begin{figure}[!htb]
\begin{center}
%\textbf{\includegraphics[width=.5\textwidth]{CE11.pdf} }
\textbf{\includegraphics[width=.9\textwidth]{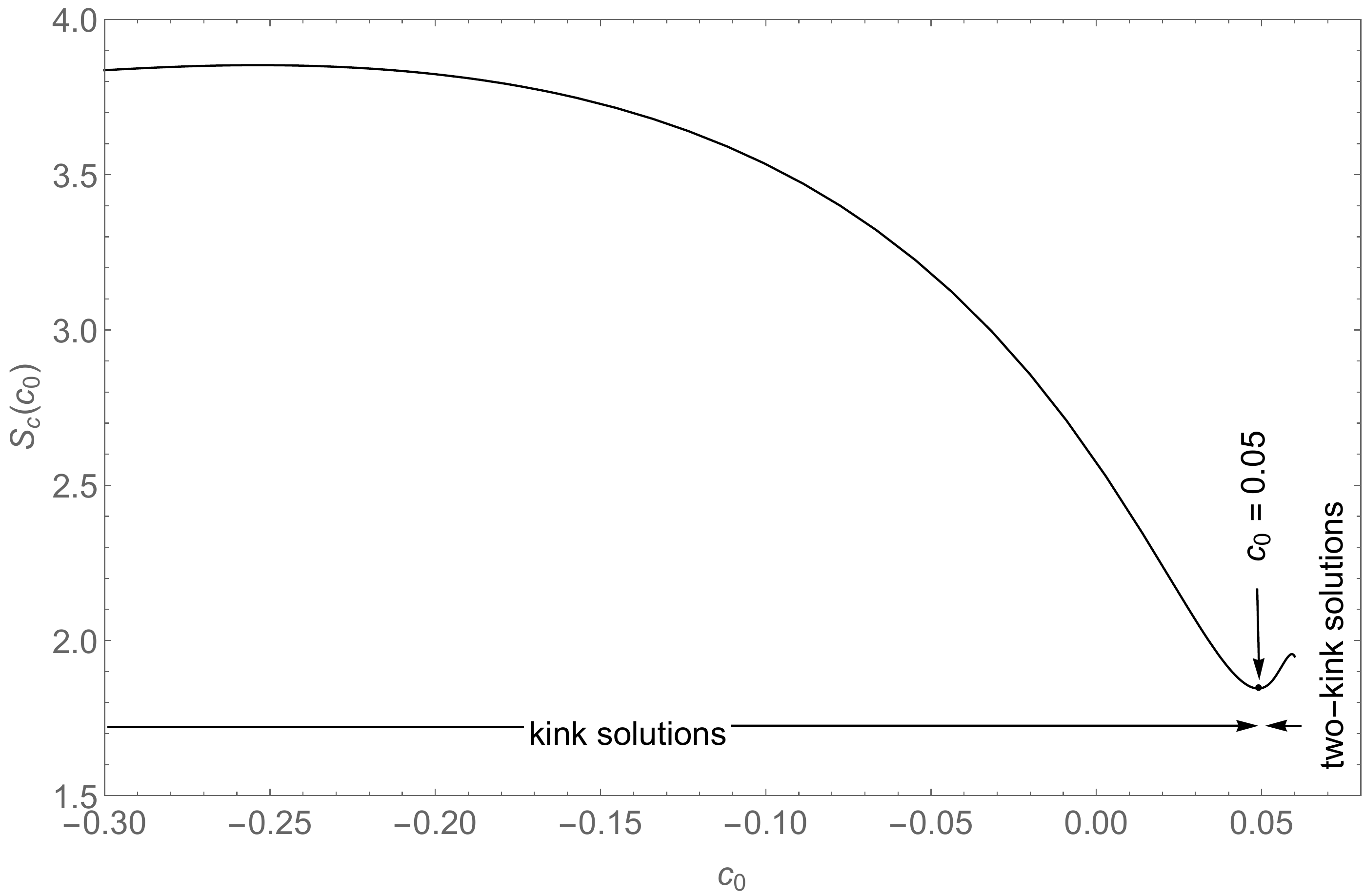} }
\end{center}
\caption{CE for DBB II solution with $a=\mu=1$}
\label{CE11}
\end{figure}

The CE for the second class of degenerate solution is
showed in Figure (\ref{CE11}). We also have a minimum entropy point defining
the frontier between the regions with kink and two-kink solutions. Since that the Fig. (\ref{fig-energy-bloch-1}) shows a more narrow and localised energy density, we verify by its CE in Fig. (\ref{CE11}) that the interval where we have two-kink solutions should be
very small. The minimum entropy $%
c_0=0.05$ corresponds to the beginning of the formation of two-kink
solutions. This also can be verified by the scalar field and its derivative
in terms of $c_0$ near the phase transition, as plotted in Figure (\ref%
{CE10}). For the kink structure, the $\phi^{\prime }(0)$ must have a
maximum. However, in the interval $0.05<c_0<0.06$ we have a local minimum to 
$\phi^{\prime }(0)$ indicating the emergence of two-kink profile.

\begin{figure}[!htb]
\begin{center}
%\textbf{\includegraphics[width=.54\textwidth]{CE10.pdf} }
\textbf{\includegraphics[width=.9\textwidth]{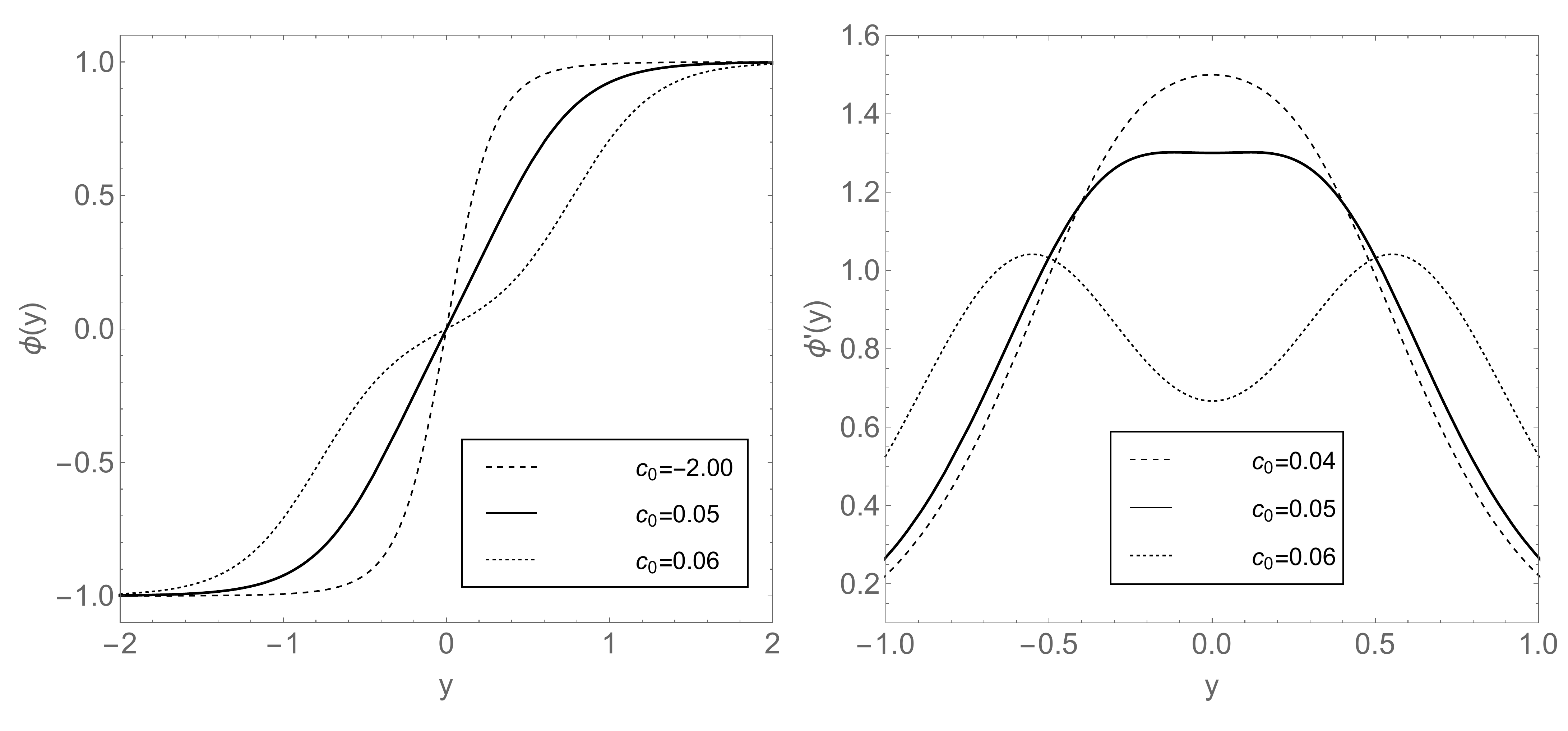} }
\end{center}
\caption{Scalar field solution for DBB II case (left) and its first
derivative (right) for $a=\mu=1$.}
\label{CE10}
\end{figure}

\section{\textbf{Discussion and conclusions}}

We investigated the properties of a $5D$ braneworld generated
by two scalar fields coupled to gravity from the viewpoint of the Configurational Entropy (CE) measure. The Bloch brane model is especially interesting in this scenario because it has a
degenerate spatially-localised energy. The connection with the entropic
information and the model is stated via the
matter-energy density along the extra dimension. From the Fourier transform
of the energy density, we construct a relation between each
degenerate energy density and its resulting CE.
%we calculate the modal fraction in terms of the degenerate parameter $c_0$. Therefore, 

The first new result is revealed when we consider the basic two-field thick
brane setup. The increasing of the brane thickness and the raising of the
internal structure occurs at the lowest entropy values. There is no phase
transition of the scalar field solutions of the basic Bloch brane and this
is expressed with the absence of local or global minima in the CE, as showed
in Fig. (\ref{CE7}).

%Turning our attention to the degenerate cases, the CE reveal us an
%interesting connection with the brane formation depicted by the behaviour of
%the scalar field solutions at minimum entropy.

For the Degenerate I Bloch Brane (DBB I) solution, the information entropy reveals us special
details concerning the brane formation. The link between the CE and the
degenerate energy density solutions is presented in Fig. (\ref{CE9}). The
minimum CE for this case is $1.32$ and corresponds to
the degenerate parameter $c_0=-2.3$. This is the turning point of the
DBB I solution. Regarding the energy density, the lowest
CE value corresponds to the raising of the internal structure, which can be
observed in Fig. (\ref{CE2}). We also observed that at minimum $S(\tilde{f})$
there is a phase transition where the kink solution to $\phi(y)$ converges
into a two-kink one.

The DBB II case also presents a splitting effect in the energy
density, however, it happens to a distinct value of the CE at the scalar
field phase transition. The region where the CE tends to the minimum value
is narrow, which reflects the narrow density energy distributions for this second model and, as expected, the interval where the two-kink solutions exists is very
small if compared with DBB I scenario.
%.In this case, there is also a specific value to the degenerate parameter where the kink solutions transit to the two-kink structure. 

One very important consequence of our analysis is that the flat region in the warp factor undergo a kind of phase transition in a certain value of the degeneracy parameter, which is designated by the CE. Thus, such approach enables to predict which is the approximate value of the degeneracy parameter, and if  the confining mechanism for the bulk particles will occur in that internal region. For instance, in scenarios where there is the localisation of fermions \cite{degenerate-Bloch-Brane-2} on the degenerate Bloch branes, the CE provides the correct values of $c_0$ for the localisation of fermionic zero modes inside the branes. Here, it is important to remark that the values of  $c_0$ are in accordance with that one found in ref. \cite{degenerate-Bloch-Brane-2}. Moreover, for the DBB II model, the double-kink region enables the existence of a strong resonant graviton coupled to brane with a larger lifetime \cite{PLBcruz2014}.

The investigation in braneworld models using the CE
approach can reveal interesting features of these kind of models. We will continue
addressing this issue in future works.
%Configurational Entropy

\section{\textbf{Acknowledgments}}

This work was partially supported by the Brazilian agencies Coordena\c{c}%
\~ao de Aperfei\c{c}oamento de Pessoal de N\'ivel Superior (CAPES) (grant
no. 99999.\newline
006822/2014-02), Conselho Nacional de Desenvolvimento Cient\'{\i}fico e
Tecnol\'{o}gico (CNPq) (grant numbers 305766/2012-0 and 305678/2015-9) and
S\~ao Paulo Research Foundation (FAPESP) (grant 2016/03276-5). The authors are very thankful for the referee suggestions.

%\textbf{The authors are very thankful to the Referee for pointing the need to better motivate the approach used for obtaining CE on the degenerate Bloch branes.} 

\end{document}